\documentclass{article}
 \newcommand{\ben}{\begin{enumerate}}
\newcommand{\een}{\end{enumerate}}
\newcommand{\beq}{\begin{equation}}
\newcommand{\eeq}{\end{equation}}
\newcommand{\bse}{\begin{subequation}}
\newcommand{\ese}{\end{subequation}}
\newcommand{\bea}{\begin{eqnarray}}
\newcommand{\eea}{\end{eqnarray}}
\newcommand{\bc}{\begin{center}}
\newcommand{\ec}{\end{center}}

\begin{document}

\begin{center}
{\Large\bf 16D anisotropic  inharmonic  oscillator and  \\  [2mm] 9D related
(MICZ-)Kepler-like systems}\\[3mm]
{\bf A. Lavrenov} \\[3mm]
Belarusian  State Pedagogical  University,  Minsk, Belarus\\
18, Sovetskaya St., Minsk, Belarus, 220050\\
e-mail: lavrenov@bspu.by\\[3mm]

{\bf I. Lavrenov} \\[3mm]
Octonion technology Ltd,  Minsk, Belarus\\
25, Y.Kupaly Str., Office 407, Minsk, Belarus, 220030.\\
e-mail: lanin99@mail.ru
\end{center}

\vspace{3mm}
\begin{abstract}
We present some generalization of 16D oscillator by  anisotropic  and nonlinear inharmonic terms and its dual analog for 9D related MICZ-Kepler systems by generalized version of the Kustaanheimo-Stiefel transformation. 
The solvability of the Schr\"{o}dinger equation of the these problems by the variables separation method are discussed in different coordinates.
\end{abstract}


\section{Introduction}
\thispagestyle{empty}

The oscillator and Kepler systems are the best known
examples  of exactly solvable tasks in few coordinate systems \cite{perelomov}.
They are dual related with each other through the Hurwitz transformations  in the dimensions of it's spaces, realizing the Hopf bundles. All favorite properties for given systems exist due to it's hidden symmetry. A few deformations of oscillator and Kepler systems can be realized by the appropriate reduction procedures of the initial hidden symmetry. 
There are anisotropic oscillator, nonlinear inharmonic oscillator,  Kepler  system with additional linear potential (its relevance to the Stark effect),  two-center Kepler system \cite{perelomov} and so on \cite{kno}. The enumerous literature is devoted to this topic  and has already developed academic logic and style of presentation of the materials for this topic (for example \cite{no} - \cite{van0}). 

On the other hand, in a number of their works, the authors \cite{van1} - \cite{van6} thoroughly worked out many questions of the last case of the Hopf bundle associated with 16-dimensional harmonic oscillator and 9-dimensional Coulomb problems. However, some moments turned out to be not considered by them and will be analyzed in this work. Thus, the goal of this paper is to give some generalization of 16D oscillator by  anisotropic  and nonlinear inharmonic terms and its dual analog for 9D related MICZ-Kepler systems by generalized version of the Kustaanheimo-Stiefel transformation. 

The outline is as follows. As already mentioned first,  we give the necessary information on result of the cycle of works by the authors of \cite{van1} - \cite{van6} in Section 2. 
Our main result will be the object of Sections 3 - 5 where the dual connection between  the some deformations for the MICZ-Kepler system and 16-dimensional oscillator will be analyzed and shown. The exact analytical solutions of the Schrödinger equation for abovementioned  problems are discused for a few coordinates systems. 
Brief concluding remarks will follow. 


\section{Dual connection between  the 9D MICZ-Kepler and a 16D isotropic harmonic oscillator problems}

 Let us consider a 16 dimensional model, the Hamiltonian which is the sum of the two  independent Hamiltonians of dimension 8, each with its own potential.
\begin{eqnarray}
H=H_1+H_2=\sum_{1\le a\le 2}H_a
=
\sum_{1\le a\le 2}
\left[
-\frac{1}{2}
\frac{\partial^2}{\partial x_{s} \partial x_{s}} + V( x_{s} x_{s}) 
\right]=\nonumber\\
=
\left[
-\frac{1}{2}
\frac{\partial^2}{\partial u_{s} \partial u_{s}} + V( u_{s} u_{s}) 
\right]
+ 
\left[
-\frac{1}{2}
\frac{\partial^2}{\partial v_{s} \partial v_{s}} + V( v_{s} v_{s}) 
\right],
\label{1}
\end{eqnarray}
where $u_s, v_t\,\,(s, t=1,\ldots, 8)$ are Cartesian coordinates of the space
${\rm I \!R}^{16}$. 
Here and futher on, 
1) the small Greek
letters $\lambda$, etc run from $1$ to $9$; 2) the small
Latin letters $j, k, s, t$, etc run from $1$ to $8$; 3) use the
Einstein convention: the repeated index is always summed up, if unless stated otherwise.

Further, we will keep in mind the independence of these 8D Hamiltonians from each other, which provide an anisotropic effect in our problem.
If we choose a quadratic potential $V_{ho}$ and one value of the oscillation frequency $\omega$ everywhere 
\begin{eqnarray}
 V_{ho} (x_{s} x_{s}) =
 \frac {\omega_a^ 2 x_{s} x_{s}} {2}; \label{2} \\
\omega= \omega_1= \omega_2, \label{3}
\end{eqnarray}
 then we obtain the model of a 16-dimensional isotropic harmonic oscillator  $\cite{van1}$
\begin{eqnarray}
\hat { H_{0}}  \psi({\bf u, v})=
 \left[-\frac{1}{2}(
\frac{{\partial}^2}{\partial u_{s} \partial u_{s}} +
\frac{{\partial}^2}{\partial v_{s} \partial v_{s}}) +
\frac{ \omega^2}{2}
\left( u_{s}  u_{s}+v_{s}  v_{s}
\right)\right] \psi = Z  \psi,
\label{4}
\end{eqnarray}
Here  $ \omega_a, \ Z_a \ (Z=Z_1+Z_2)$  have positive real values and are correspondingly the frequency and energy of the each harmonic oscillator.

According to  \cite{van0}, the Hurwitz transformation ${\rm I \!R}^{16} \to {\rm I \!R}^9$  that connects the 9D real space ${\rm I \!R}^9$  of Cartesian coordinates  $ x_{1}, x_{2},\ldots,x_{9}$    to the 16D real space ${\rm I \!R}^{16}$  of Cartesian coordinates $ u_{1}, u_{2},\ldots, u_{8}, v_{1}, v_{2},\ldots, v_{8}$  can be written as follows:
 \bea
 x_{k}\rightarrow 2 (\Gamma_{k})_{st} u_s v_t\;,    \quad  \quad
 x_{9}\rightarrow  u_{s}  u_{s} - v_{s}  v_{s} ,
  ,\label{5} \eea

After the application of the Hurwitz transformation (\ref{5})  to equation (\ref{4})  the Schr\"{o}dinger equation for the nine-dimensional MIC-Kepler problem in atomic units (thus $m= e = \hbar=1$)   has the form \cite{van4}
\bea
\hat { H'_{0}}  \psi({\bf r})=
\left(
\frac{1}{2}
{\hat \pi}^2+
\frac{{\hat Q}^2}{8 r^2}-
\frac{Z}{r}
\right)\psi= E \psi
,\label{6} \eea
in which $Z$  is the nuclear charge of the Coulomb interaction, $E$  is the energy of the particle, $ {\hat \pi}^2={\hat \pi}_{\lambda} {\hat \pi}_{\lambda}$  with impulse operators  are determined  by
\bea
{\hat \pi}_{j}=-i
\frac{{\partial}}{\partial  x_{j}} +
A_{k} {\hat Q}_{kj}, \
{\hat \pi}_{9}=-i\frac{{\partial}}{\partial  x_{9}}
,\label{7} \eea
where the potential vectors $  A_{k}=\frac{x_{k}}{r(r + x_{9})}$,  squared operator
 $ {\hat Q}^2={\hat Q}_{jk} {\hat Q}_{jk}$ and
$  \qquad r= \sqrt{ x_{\lambda} x_{\lambda}} = u_{s}  u_{s} + v_{s}  v_{s}$  is distance in nine-dimensional space.

By plugging formulae (\ref{7}) into Eq. (\ref{6}) , the Schr\"odinger equation for $\hat {H'_0}$ now becomes according to works  \cite{van5} - \cite{van6}
\begin{eqnarray}
\hat { H'_{0}} \ \psi=
\left[
- \frac{\Delta}{2}+
\frac{ Q_{kj} L_{kj} }{2 r (r+x_{9}) }+
\frac{ Q^2} {4 r (r+x_{9}) }-
\frac{Z}{r}
\right] \psi({\bf r})= E  \psi({\bf r}).
\label{8}
\end{eqnarray}

We notify that in Eqs. (\ref{4}) and (\ref{6}) the roles of $ E= - \frac{\omega^2}{2}$ and $Z$  are interchanged. The variables $E$ and $Z$ become  a negative number that denotes the energy of bound states and a parameter defining the `charge'  value in the Coulomb potential respectively. 

Thus, in the paper \cite{van1} it is shown that a 16-dimensional isotropic harmonic oscillator and the nine-dimensional MICZ-Kepler problem described by the Schr\"{o}dinger equation  in Eqs. (\ref{4}) and (\ref{6})  are dual.


\section{16D anisotropic  and nonlinear inharmonic oscillator}

In a work \cite{lan1} we proposed next generalization of the Schr\"odinger equation for  so-called a 16D double singular oscillator:
\begin{eqnarray}
\hat {H} \ \psi({\bf u, v})=
\left[{\hat H}_0+
\frac{c_1}{u^2_1+...+u^2_8}+
\frac{c_2}{v^2_{1}+...+v^2_8}
\right] \psi = Z  \psi,
\label{9}
\end{eqnarray}
where $c_{1}, c_{2}$ nonnegative constants; ${\hat H}_0$ is Hamiltonian  of a 16-dimensional isotropic harmonic oscillator determined earlier (\ref{4}).

In other words, instead of harmonic oscillator potential $V_{ho}$  in Eqs. (\ref{1}), we chose the potential of a singular oscillator
\begin{eqnarray}
V_{sho}( x_{s} x_{s}) =
V_{ho}( x_{s} x_{s}) +
\frac{c_a}{x_{s} x_{s}} =
\frac{ \omega^2  x_{s} x_{s}}{2}
+
\frac{c_a}{x_{s} x_{s}}.
\label{10}
\end{eqnarray}

To make variables separation in each 8D real space ${\rm I \!R}^8$ where   $ x_{1}, x_{2},\ldots,x_{8}$ are Cartesian coordinates, we have introduced  it's hyperspherical coordinates: ($\phi_{1},\dots, \phi_{7}$)  are the hyperspherical angles and $r$ is the hyperradius by
\begin{eqnarray}
\nonumber&  x_{8}&=r\cos(\phi_{7}),
\nonumber\\&  x_{7}&=r\sin(\phi_{7})\cos(\phi_{6}),
\nonumber\\&...&
\nonumber\\&...&
\nonumber\\& x_{2}&=r \sin(\phi_{7})\sin(\phi_{6})\cdots \sin(\phi_{2}) \cos(\phi_{1}),
\nonumber\\& x_{1}&=r \sin(\phi_{7})\sin(\phi_{6})\cdots \sin(\phi_{2}) \sin(\phi_{1}),
\label{11}
\end{eqnarray}

Using the ansatz $ \Psi({\bf r})=\Psi(r,\phi_{1},\cdots,\phi_{7})=R(r) \Omega(\phi)$
the Schr\"odinger equation for  each $H_a$ can be rewritten in terms of it's separated equations of hyperradius $r$ and angular variables $\phi$ as follows:
\begin{eqnarray}
\left[
\frac{1}{r^{7}}\frac{\partial}{\partial r} ( r^{7} \frac{\partial}{ \partial r})+
\left(
2 Z_a-\omega^2 r^2
\right)-
\frac{\Lambda^2+2c_a}{r^2}
\right]R (r)=0,\label{12} \\
\left[
\Lambda^2-L(L+6)
\right]
\Omega(\phi)=0,
\label{13}
\end{eqnarray}
where $Z_a $ are the eigenvalues of $H_a$, but $Z_1+Z_2=Z;  \quad \Lambda=L(L+6)$ is the separation constant and is also an eigenvalue of the operator $\Lambda^2$ ($\ref{13}$).

A solution  of Eq. ($\ref{12}$) 
are given in terms of a special function 
\begin{eqnarray}
R_{N L}(r) = C_{NL}  r^{L'} e^{-\frac{ \omega r^2}{2}} {}_1F_1 (-N; L'+4; \omega r^2) \label{14}
\end{eqnarray}
where
$L'(L'+6) =L(L+6) + 2 c_{a}.$

Thus, it can be said that, firstly, the actual multidimensional problem was reduced
to one-dimensional case for hyperradius $r$, and, secondly, our proposed generalization according to the above formulas is reduced to additive terms in the negative power of $r$ for the initial Hamiltonians. 

On the other hand, recall, that there are quasi-exactly problems which occupy an intermediate place between exactly solvable problems and non-solvable ones. The theory of  quasi-exactly systems (QES) gives 
the next generalization or the family of potentials in the direction of degrees $ r $ less than 2 for a fixed $N$, by
\begin{equation}
\label{15}
 V_{<2}(r)\ =V_{sho} + a r + \frac{b}{r}\  =
 \ \frac {\omega^2 r^2}{2}  + \frac{c}{r^2} + a r + \frac{b}{r}\
\end{equation}
with the eigenfunctions 
\begin{equation}
\label{16}
        R (r) \ = \  p_{N-1} (r) r^{l'-c'}  e^{-\frac{b'}{2}\, r^2  - a'\, r}\ ,
\end{equation}
where there are the following reassignment of constants from work \cite{Turbiner1} to our designation $\omega^2=2 b'^2; \quad a =2 a' b';  \quad b=-a' (D - 2c');  \quad c=c'(c'-D+1); \quad d=a'^2 - b'(2 N+D -1-2c'); \quad D=d'+2l'-1$; \quad $p_{N-1}(r)$ are polynomial of the ($N-1$)-th degree.

This QES potential appears in a number of applications to the systems with two electrons (\cite{Turbiner2} -  \cite{Turbiner3}).

Also at present we want to go to oppositive direction and to consider the potential which depenfing on $u, v$ for it's degrees more than 2. The QES theory gives the other generalization in this direction or the family of potentials for a fixed $N$, by
see \cite{Turbiner1},
\begin{equation}
\label{17}
      V_{>2}(z) =V_{sho} + b r^4 + a r^6   \ =
        (\omega r)^2/2 + \frac{c}{r^{2}} + b r^4 + a r^6   \ ,
\end{equation}
with the eigenfunctions 
\begin{equation}
\label{18}
        R (r) \ = \  p_{N-1} (r^2) r^{l'-c'}  e ^ {-\frac{a'r^4}{4} - \frac{b'r^2}{2}},
\end{equation}
where the following reassignment of constants from work \cite{Turbiner1}  to our designation $\omega^2=2[b'^2 - (4 N+D - 2c'-1)a']; \quad a = a'^2;  \quad b=2 a' b';  \quad c= c'(c' - D+1)$.

The one-dimensional Hamiltonian of the nonrelativistic quantum systems with this anharmonic potential $(\ref{17})$ is well known as the crucial example that stimulated the investigation of quasi-exactly solvable systems.

Thus, we offer four different models of the 16D anisotropic and nonlinear anharmonic oscillator in QES class. Each model is represented by a sum of two independent oscillators of dimensions $ D = 8 $ with various anisotropic and nonlinear anharmonic terms of the potential.
In other words, we will further consider the following Hamiltonians of dimension $ D = 8 $:
\begin{eqnarray}
_{<2}H = &\left[-\frac{1}{2}
\frac{\partial^2}{\partial x_{s} \partial x_{s}} +  V_{<2}( x_{s} x_{s}) \right];\nonumber\\
V_{<2}(x_{s} x_{s})\ = & V_{sho}( x_{s} x_{s}) + \frac{b}{ \sqrt{ x_{s} x_{s} }}
 + a \sqrt { x_{s} x_{s} }  \nonumber\\
   = &\ \frac {\omega^2  x_{s} x_{s}}{2}  + \frac{c}{ x_{s} x_{s}}  + \frac{b}{ \sqrt{ x_{s} x_{s} }}
 + a \sqrt { x_{s} x_{s} } \  \label{19}
\end{eqnarray}
and
\begin{eqnarray}
_{>2}H = &\left[-\frac{1}{2}
\frac{\partial^2}{\partial x_{s} \partial x_{s}} + 
 V_{>2}( x_{s} x_{s})
\right];\nonumber \\
V_{>2}(x_{s} x_{s})\ = &V_{sho}( x_{s} x_{s}) +
 b (x_{s} x_{s})^{2} + a (x_{s} x_{s})^{3} \nonumber\\
  = &\ \frac {\omega^2  x_{s} x_{s}}{2}  + \frac{c}{ x_{s} x_{s}} + b (x_{s} x_{s})^{2} + a (x_{s} x_{s})^{3}. \label{20}
\end{eqnarray}

In hyperspherical coordinates, the potentials of these Hamiltonians are successively reduced to the potential ($ \ref{15} $) and ($ \ref{17}$).
In any case, the final wave function $  \psi({\bf u, v})$  of Eq. ($\ref{1}$) will be represented by the product of the wave functions of each oscillator $\Psi({\bf r_a}) \equiv  R (r_a) \Omega(\phi_a)$ of Eq. ($\ref{19}$) -  ($\ref{20}$):
\begin{itemize}
 \item  { 
Model 1
\begin{eqnarray}
H \psi_1({\bf u, v})& =& \left[{}_{<2}H_1 + {}_{<2}H_2\right]  \psi_{<2}({\bf u}) \psi_{<2}({\bf v})    \nonumber \\
& \equiv &\left[{}_{<2}H_1+{}_{<2}H_2\right]  \Psi_{<2}({\bf r_1}) \Psi_{<2}({\bf r_2})  \nonumber    \\
&   =  &  _{<2}H_1\left[\Psi_{<2}({\bf r_1}) \Psi_{<2}({\bf r_2})\right] +
  _{<2}H_2\left[\Psi_{<2}({\bf r_1}) \Psi_{<2}({\bf r_2})\right]  \nonumber\\
& =  &  \Psi_{<2}({\bf r_2}) \left[ _{<2}H_1\Psi_{<2}({\bf r_1}) \right] +
\Psi_{<2}({\bf r_1}) \left[ _{<2}H_2\Psi_{<2}({\bf r_2}) \right]  \nonumber\\
&  =  &   \Psi_{<2} ({\bf r_2})  \left[ Z_1\Psi_{<2}({\bf r_1}) \right] +
\Psi_{<2}({\bf r_1}) \left [Z_2 \Psi_{<2}({\bf r_2}) \right]  \nonumber\\
&   =  &   \left[ Z_1 + Z_2 \right]  \Psi_{<2}({\bf r_1}) \Psi_{<2}({\bf r_2})                          \equiv  Z \Psi_{<2}({\bf r_1}) \Psi_{<2}({\bf r_2})           \nonumber\\
      &   \equiv  &  Z R_{<2}({r_1}) \Omega_1(\phi_1) R_{<2}({r_2})\Omega_2(\phi_2)  =Z \psi_1({\bf u, v})  
       \label{21}
\end{eqnarray}
}
 \item 
 {
 Model 2
\begin{eqnarray}
\label{22}
H \psi_2({\bf u, v})& =& \left[{}_{<2}H_1 + {}_{>2}H_2\right]  \psi_{<2}({\bf u}) \psi_{>2}({\bf v})    \nonumber \\
& \equiv & Z \Psi_{<2}({\bf r_1}) \Psi_{>2}({\bf r_2})
\end{eqnarray}
}
 \item 
 {
 Model 3
\begin{eqnarray}
\label{23}
H \psi_3({\bf u, v})& =& \left[{}_{>2}H_1 + {}_{<2}H_2\right]  \psi_{>2}({\bf u}) \psi_{<2}({\bf v})    \nonumber \\
& \equiv & Z \Psi_{>2}({\bf r_1}) \Psi_{<2}({\bf r_2})
\end{eqnarray}
}
 \item 
 {
 Model 4
\begin{eqnarray}
\label{24}
H \psi_4({\bf u, v})& =& \left[{}_{>2}H_1 + {}_{>2}H_2\right]  \psi_{>2}({\bf u}) \psi_{>2}({\bf v})    \nonumber \\
& \equiv & Z  \Psi_{>2}({\bf r_1}) \Psi_{>2}({\bf r_2})
\end{eqnarray}
}
\end{itemize}

Recall that the hyperradius part  $R (r_a)$ of the eigenfunction $\Psi({\bf r_a})$ has the form  ($ \ref{16}$)  and ($ \ref{18} $), respectively, depending on the potential
($ \ref{15}$) and ($ \ref{17} $).


\section{9D related (MICZ-) Kepler-like systems as dual analog of 16D anisotropic and nonlinear inharmonic oscillator}

 	The Hurwitz transformation links the harmonic oscillator with the Coulomb problem.
Therefore,  9D related MICZ-Kepler systems are considered as dual analog of 16D oscillator.  After the application of the Hurwitz transformation (\ref{5})  to equation (\ref{4}) 
and by plugging formulae (\ref{7}) into Eq. (\ref{6}),
the Schr\"{o}dinger equation for the nine-dimensional MICZ-Kepler problem in atomic units (thus $m= e = \hbar=1$)  $\hat {H'_0}$   now becomes according to works  \cite{van5} - \cite{van6}
\begin{eqnarray}
\hat { H'_{0}} \ \psi=
\left[
- \frac{\Delta}{2}+
\frac{ J^2 -L^2} {4 r (r+x_{9}) } +
 V'_{C}
\right] \psi({\bf r})= E  \psi({\bf r}).
\label{25}
\end{eqnarray}
where  $ E= - \frac{\omega^2}{2},  V'_{C}=-\frac{Z}{r}$ and $J_{kj}=L_{kj}+Q_{kj}$.

This means that this transition from one equation (\ref {4}) to another (\ref {6}-\ref {8}) or
(\ref {25})  can be interpreted as some change of expression
\begin{eqnarray*}
\sum_{1\le a\le 2}
\left[
 V_{iho}( x_{s} x_{s}) -Z 
 \right] 
& \equiv
V_{iho 1}( u_{s} u_{s})+V_{iho 2}( v_{s} v_{s}) -Z_1 -Z_2 
&=\\
\sum_{1\le a\le 2}
\left[ 
\omega^2 x_{s} x_{s}/2  - Z
\right] 
& \equiv 
\omega^2 u_{s} u_{s}/2  - Z_1 + 
\omega^2 v_{s} v_{s}/2  - Z_2 
&=\\
\sum_{1\le a\le 2}
\left[
- E  x_{s} x_{s}  - Z
\right] 
& \equiv 
- E_1  u_{s} u_{s}  - Z_1 - E_2  v_{s} v_{s}  - Z_2  
& \\ 
\quad \to \quad 
V'_{JLC} - E  
& \equiv 
V'_{JL}  +
V'_{C}
- E
  \equiv 
\frac{ J^2 -L^2} {4 r (r+x_{9}) } -
\frac{Z}{r}
- E&
\end{eqnarray*}
or
\begin{eqnarray*}
\sum_{1\le a\le 2}
\left[
V_{iho}( x_{s} x_{s}) -Z 
\right] 
 =&
\sum_{1\le a\le 2}
\left[
\omega^2 x_{s} x_{s}/2  - Z 
\right] 
\equiv &
\sum_{1\le a\le 2}
\left[
 -E  x_{s} x_{s}  - Z 
 \right] 
 \\
 \quad \to \quad 
V'_{JLC} - E  =&
 \frac {1} {r} 
\left[
\frac{ J^2 -L^2} {4 (r+x_{9}) } - 
Z - E r
 \right] 
\equiv &
 \frac {1} {r} 
\left[
r V'_{JL} +
r(  V'_{C} -  E)
 \right] 
\end{eqnarray*}

Considering the additive and singular terms $c_{a}/ x_{s} x_{s}$ to the potential of a harmonic oscillator  $V_{ho}( x_{s} x_{s})$, we obtain the potential of a singular harmonic oscillator $V_{sho}( x_{s} x_{s})$  $(\ref{10})$, which after Hurwitz transformation will also have an additional additive terms to the previously obtained Coulomb potential $V'_{C}$.
Therefore,
 we get the following $9D$  generalized MIC-Kepler system with non central terms
\bea
\hat {H'} \ \psi({\bf r})=
\left[\hat { H'_{0}}+
\frac{2 c_1}{r(r+x_{9})} + 
\frac{2 c_2}{r(r-x_{9})}
\right]\psi= E \psi
,\label{26} \eea
where ${\hat H'}_0$ is Hamiltonian  of the nine-dimensional MICZ-Kepler problem determined earlier  (\ref{25}). 

This was not hard to do if is remember what is taking place:
\begin{eqnarray*}
 \qquad r= \sqrt{ x_{\lambda} x_{\lambda}} = u_{s}  u_{s} + v_{s}  v_{s}; \
 \qquad  x_{9} = u_{s}  u_{s} - v_{s}  v_{s}\\
  \qquad 2 u_{s}  u_{s} = r + x_{9};   \qquad  2  v_{s}  v_{s} = r - x_{9}.
\end{eqnarray*}

Thus, now we are ready to determine the potential of the dual analog for our generalizations  of the 16D oscillator [ 4 models  
$(\ref{21}$) - ($\ref{24}$ ] as 9D related MIC-Kepler systems in different coordinates. The solvability of the Schr\"{o}dinger equation of the these problems by the variables separation method will be discussed in spherical and parabolic coordinates.
 

\subsection{Variables separation  in spherical coordinates}

According to  to works  \cite{van5} - \cite{van6}, the Cartesian coordinates  $ x_{1}, x_{2},\ldots,x_{9}$  of 9D real space ${\rm I \!R}^9$  are defined by
the nine-dimensional spherical coordinates
\begin{eqnarray}
\nonumber&  x_{9}&=r\cos(\theta),
\nonumber\\&  x_{8}&=r\sin(\theta)\cos(\phi_{6}),
\nonumber\\&...&
\nonumber\\&...&
\nonumber\\& x_{2}&=r \sin(\theta)\sin(\phi_{6})\cdots \sin(\phi_{1}) \cos(\phi_{0}),
\nonumber\\& x_{1}&=r \sin(\theta)\sin(\phi_{6})\cdots \sin(\phi_{1}) \sin(\phi_{0}),
\label{27}
\end{eqnarray}

Given these definitions, we obtain useful relations:
\begin{eqnarray*}
  \qquad 2 u_{s}  u_{s} = r + x_{9} \equiv r(1 + \cos (\theta)) = 2 r \cos^2 (\frac {\theta}{2}) ; \\
   \qquad 2  v_{s}  v_{s} = r - x_{9} \equiv r(1 - \cos (\theta)) = 2 r \sin^2 (\frac {\theta}{2}).
\end{eqnarray*}
In other words, we received next typies of potential  in  spherical coordinates
\begin{eqnarray*}
V_{<2}(u_{s} u_{s})  &=& \frac {\omega_1^2  u_{s} u_{s}}{2}  + \frac{c_1}{ u_{s} u_{s}}  + \frac{b_1}{ \sqrt{ u_{s} u_{s} }}
 + a_1 \sqrt { u_{s} u_{s} } 
\\
 &=& \frac {\omega_1^2  r \cos^2 (\frac {\theta}{2}) }{2}  + \frac{c_1}{  r \cos^2 (\frac {\theta}{2}) }  + \frac{b_1}{ \sqrt{  r \cos^2 (\frac {\theta}{2}) ) }}
 + a_1 \sqrt { r \cos^2 (\frac {\theta}{2})  }    \\
 &=& r \left[ - E_1 \cos^2 (\frac {\theta}{2})  + \frac{c_1}{  r^2 \cos^2 (\frac {\theta}{2}) }  + \frac{b_1}{ \sqrt{  r^3  \cos^2 (\frac {\theta}{2}) }}
 + a_1 \sqrt {  \frac {\cos^2 (\frac {\theta}{2}) }{r} }  \right]  \\
  V_{<2}(v_{s} v_{s})&=& r \left[ - E_2 \sin^2 (\frac {\theta}{2})  + \frac{c_2}{  r^2 \sin^2 (\frac {\theta}{2}) }  + \frac{b_2}{ \sqrt{  r^3  \sin^2 (\frac {\theta}{2}) }}
 + a_2\sqrt {  \frac {\sin^2 (\frac {\theta}{2}) }{r} }  \right]  
 \end{eqnarray*}
and
\begin{eqnarray*}
 V_{>2}(u_{s} u_{s}) &=& \frac {\omega_1^2  u_{s} u_{s}}{2}  + \frac{c_1}{ u_{s} u_{s}}  +
 b_1 (u_{s} u_{s})^{2} + a_1 (u_{s} u_{s})^{3}  \\
   &=&\frac {\omega_1^2  r \cos^2 (\frac {\theta}{2}) }{2}  + \frac{c_1}{  r \cos^2 (\frac {\theta}{2}) }  +
    b_1 (r \cos^2 (\frac {\theta}{2}))^{2} + a_1 (r \cos^2 (\frac {\theta}{2}))^{3}   \\
  &=& r \left[ - E_1 \cos^2 (\frac {\theta}{2})  + \frac{c_1}{  r^2 \cos^2 (\frac {\theta}{2}) }   +b_1 r \cos^4 (\frac {\theta}{2}) + a_1 r^2 \cos^6 (\frac {\theta}{2})  \right]  \\
   V_{>2}(v_{s} v_{s}) & =& 
   r \left[ - E_2 \sin^2 (\frac {\theta}{2})  + \frac{c_2}{  r^2 \sin^2 (\frac {\theta}{2}) }   +b_2 r \sin^4 (\frac {\theta}{2}) + a_2 r^2 \sin^6 (\frac {\theta}{2})  \right] 
  \end{eqnarray*}

This means that all additive terms to the Coulomb potential have a double dependence both on the hyperradius $r$ and on the angular variable $\theta$. It is reasonable to assume that: 1) their separability is possible only when the entire dependence on the angular variable $\theta$ is collected with one functional dependence on the hyperradius $r$, and in other cases it is absent; 2) the type of functional dependence on hyperradius  $r$ is determined by the existing functional dependence on hyperradius  $r$ near derivatives with respect to the angular variable $\frac{\partial}{ \partial \theta}$.

Following this logic leads to the scheme of separation of variables for the (generalized) Coulomb case with the amendment in the equation for the radial variable, where additional terms appear without angular dependence, i.e. instead of $ -r (V '_ {C} - E) = Z + E r $ we get $ W' (r) $.
Thus,
the Schr\"odinger equation of our generalized MICZ-Kepler system with new terms rewritten in terms of next separated equations of variables $r, \theta$ and angular variables $\phi, \phi'$
\begin{eqnarray}
\left[
\frac{1}{r^{8}}\frac{\partial}{\partial r} ( r^{8} \frac{\partial}{ \partial r})-
\frac{2}{r}
 W'(r)
- \frac{\Lambda}{r^2}
\right]R (r)=0,\label{28} \\
\left[
\frac{1}{\sin^{7}\theta}\frac{\partial}{ \partial \theta}( \sin^{7}\theta\frac{\partial}{ \partial \theta})  -
\frac{L(L+6)+8 c_{2}}{4 \sin^{2}\frac{\theta}{2}}-
\frac{J(J+6)+8 c_{1}}{4 \cos^{2}\frac{\theta}{2}}+ \Lambda
\right]
\Theta(\theta)=0,\label{29} \\
\left[
J^2 -J(J+6)
\right]
\Phi(\phi, \phi')=0,
\label{30}\\
\left[
L^2 -L(L+6)
\right]
\Phi(\phi, \phi')=0,
\label{31}
\end{eqnarray}
where $\Lambda=\lambda (\lambda+7)$ is the separation constant and is also an eigenvalue of the operator ($\ref{29}$).

Let us specify the value of  $W'(r) \equiv  
 \frac {1} {r} 
\left[
V(u_{s} u_{s}) + V(v_{s} v_{s}) -Z_1 - Z_2
\right]
$ 
 for each model and  the conditions on coefficients of potential for (quasi) exact solvability according to above logic.
\begin{itemize}
 \item  { 
Model 1
\begin{eqnarray*}
H \psi_1({\bf u, v}) = & \left[{}_{<2}H_1 + {}_{<2}H_2\right]  \psi_{<2}({\bf u}) \psi_{<2}({\bf v})         \equiv    Z \psi_1({\bf u, v})   \\
  W'(r) =&
   \left[
  - E_1 \cos^2 (\frac {\theta}{2})  + 
   \frac{b_1}{ \sqrt{  r^3  \cos^2 (\frac {\theta}{2}) }}
 + a_1 \sqrt {  \frac {\cos^2 (\frac {\theta}{2}) }{r} }  \right]  -  Z_1 
 \\
+& \left[ - E_2 \sin^2 (\frac {\theta}{2})  +
   \frac{b_2}{ \sqrt{  r^3  \sin^2 (\frac {\theta}{2}) }}
 + a_2\sqrt {  \frac {\sin^2 (\frac {\theta}{2}) }{r} }  \right]   - Z_2
    \end{eqnarray*}
or $E_1 =E_2; \ b_1= a_1 =b_2 = a_2=0 $.
}
 \item 
 {
 Model 2
\begin{eqnarray*}
H \psi_2({\bf u, v}) =& \left[{}_{<2}H_1 + {}_{>2}H_2\right]  \psi_{<2}({\bf u}) \psi_{>2}({\bf v})   
\equiv  Z \Psi_{<2}({\bf r_1}) \Psi_{>2}({\bf r_2})
 \\
  W'(r) =&
   \left[
  - E_1 \cos^2 (\frac {\theta}{2})  + 
   \frac{b_1}{ \sqrt{  r^3  \cos^2 (\frac {\theta}{2}) }}
 + a_1 \sqrt {  \frac {\cos^2 (\frac {\theta}{2}) }{r} }  \right]  -  Z_1 
 \\
+&
 \left[ - E_2 \sin^2 (\frac {\theta}{2})  +b_2 r \sin^4 (\frac {\theta}{2}) + a_2 r^2 \sin^6 (\frac {\theta}{2})  \right] 
         -  Z_2
  \end{eqnarray*}
  or $E_1 =E_2; \ b_1= a_1 =b_2 = a_2=0 $.
}
\item 
 {
 Model 3
\begin{eqnarray*}
H \psi_3({\bf u, v})& = \left[{}_{>2}H_1 + {}_{<2}H_2\right]  \psi_{>2}({\bf u}) \psi_{<2}({\bf v})   
\equiv  Z \Psi_{>2}({\bf r_1}) \Psi_{<2}({\bf r_2})
 \\
  W'(r) =&
   \left[
  - E_1 \cos^2 (\frac {\theta}{2})  +b_1 r \cos^4 (\frac {\theta}{2}) + a_1 r^2 \cos^6 (\frac {\theta}{2})  \right]  
  -  Z_1 
 \\
+ &
 \left[ - E_2 \sin^2 (\frac {\theta}{2})  +
   \frac{b_2}{ \sqrt{  r^3  \sin^2 (\frac {\theta}{2}) }}
 + a_2\sqrt {  \frac {\sin^2 (\frac {\theta}{2}) }{r} }  \right] 
   - Z_2
   \end{eqnarray*}
or $E_1 =E_2; \ b_1= a_1 =b_2 = a_2=0 $.
}
 \item 
 {
 Model 4
\begin{eqnarray*}
H \psi_4({\bf u, v})& = \left[{}_{>2}H_1 + {}_{>2}H_2\right]  \psi_{>2}({\bf u}) \psi_{>2}({\bf v})  
\equiv  Z  \Psi_{>2}({\bf r_1}) \Psi_{>2}({\bf r_2})
 \\
  W'(r) =&
 \left[ - E_1 \cos^2 (\frac {\theta}{2})  + b_1 r \cos^4 (\frac {\theta}{2}) + a_1 r^2 \cos^6 (\frac {\theta}{2})  \right] 
   -  Z_1 
  \\
  +&
 \left[ - E_2 \sin^2 (\frac {\theta}{2})  + b_2 r \sin^4 (\frac {\theta}{2}) + a_2 r^2 \sin^6 (\frac {\theta}{2})  \right] 
 - Z_2
       \end{eqnarray*}
   or $E_1 =E_2; \ b_1= a_1 =b_2 = a_2=0 $. 
}
\end{itemize}

Thus, we have obtained that in spherical coordinates the maximum (quasi) exactly solvable model for our proposal is only the (generalized) Coulomb model  ($\ref{26}$).


\subsection{Variables separation  in parabolic coordinates}

According to works  \cite{van5} - \cite{van6}, the Cartesian coordinates  $ x_{1}, x_{2},\ldots,x_{9}$  of 9D real space ${\rm I \!R}^9$  are defined by
the parabolic coordinates  on the $S_7$ sphere 
\begin{eqnarray}
&x_{9}&=\frac{u-v}{2},
\nonumber\\& x_{8}&=\sqrt{u v} \cos(\phi_{6}),
\nonumber\\&...&
\nonumber\\&...&
\nonumber\\&  x_{2}&=\sqrt{u v} \sin(\phi_{6})\cdots \cos( \phi_{0}),
\nonumber\\& x_{1}&=\sqrt{u v} \sin(\phi_{6})\cdots \sin( \phi_{0}),
\nonumber\\& r&= \frac{u +v}{2}.
\label{32}
\end{eqnarray}

The Schrodinger equation $H'\Psi=E\Psi$  ($\ref{25}$) is same, but the Laplace-Beltrami operator $\Delta$ in the hyperparabolic coordinates has the next form:
\begin{eqnarray*}
&&\Delta=
\frac{4}{u+v}
\left\{
\frac{1}{u^{3}}\frac{\partial}{\partial u} ( u^{4} \frac{\partial}{ \partial u})+
\frac{1}{v^{3}}\frac{\partial}{\partial r} ( v^{4} \frac{\partial}{ \partial v})
\right\}-
\frac{L^2}{u v}
\end{eqnarray*}

Given these definitions, we obtain useful relations:
\begin{eqnarray*}
 \qquad r= \sqrt{ x_{\lambda} x_{\lambda}} = u_{s}  u_{s} + v_{s}  v_{s}; \
 \qquad  x_{9} = u_{s}  u_{s} - v_{s}  v_{s}\\
     \qquad 2 u_{s}  u_{s} = r + x_{9} \equiv u ; \ 
   \qquad 2  v_{s}  v_{s} = r - x_{9} \equiv v 
\end{eqnarray*}

In other words, we received next typies of potential  in  the parabolic coordinates  
\begin{eqnarray*}
V_{<2}(u_{s} u_{s})  &=& \frac {\omega_1^2  u_{s}u_{s}}{2}  + \frac {c_1}{ u_{s}u_{s}}  + \frac {b_1}{ \sqrt{ u_{s} u_{s} }} + a_1 \sqrt { u_{s} u_{s} }   \\
&=& \frac {- u E_1}{2}  + \frac {2c_1}{u}  + b_1 \sqrt{ \frac {2}{u} }
 + a_1 \sqrt { \frac {u}{2}  }  \\
V_{<2}(v_{s} v_{s})&=&\frac {- v E_2}{2}  + \frac {2c_2}{v}  + b_2 \sqrt{ \frac {2}{v} }
 + a_2 \sqrt { \frac {v}{2}  } 
 \end{eqnarray*}
and
\begin{eqnarray*}
 V_{>2}(u_{s} u_{s}) &=& \frac {\omega_1^2  u_{s} u_{s}}{2}  + \frac{c_1}{ u_{s} u_{s}}  +  b_1 (u_{s} u_{s})^{2} + a_1 (u_{s} x_{s})^{3} \\
 &=& \frac {- u E_1}{2}  + \frac {2c_1}{u} +
  \frac {b_1 u^2}{4} + \frac {a_1 u^3}{8}   \\
V_{>2}(v_{s} v_{s}) & =& 
\frac {- u E_2}{2}  + \frac {2c_2}{u}  + 
  \frac {b_2 u^2}{4} + \frac {a_2 u^3}{8}
\end{eqnarray*}

As in the case of spherical coordinates, we have additional additive terms, but each of which now depends on only one variable. This fact makes it easy to separate variables according to the scheme of separation of variables for the (generalized) Coulomb case.

Thus, the Schr\"odinger equation of our generalized MIC-Kepler system with new terms in the hyperparabolic coordinates can be rewritten in terms of separated equations of variables $u, v$ and angular variables $\phi, \phi'$
\begin{eqnarray}
\left[
\frac{1}{u^{3}}\frac{\partial}{\partial u} ( u^{4} \frac{\partial}{ \partial u}) -
\frac{J (J+6)+8 c_{1}}{4 u} -
W'_{u}(r) 
-P
\right]U (u)=0,\label{33} \\
\left[
\frac{1}{v^{3}}\frac{\partial}{\partial v} ( v^{4} \frac{\partial}{ \partial v}) -
\frac{L (L+6)+8 c_{2}}{4 v} -
W'_{v}(r) 
+P
\right]V (v)=0,\label{34} \\
\left[
J^2 -J(J+6)
\right]
\Phi(\phi, \phi')=0,
\label{35}\\
\left[
L^2 -L(L+6)
\right]
\Phi(\phi, \phi')=0,
\label{36}
\end{eqnarray}
where $P$ is the separation constant.

Let us specify the value of  
$
W'_{u}(r) 
\equiv  
\left[
V(u_{s} u_{s})  -Z_1 
\right]
$ 
 and 
 $
 W'_{v}(r) 
 \equiv  
\left[
V(v_{s} v_{s})  - Z_2
\right]
$ 
for each model
\begin{itemize}
 \item  { 
Model 1
\begin{eqnarray}
\label{37}
H \psi_1({\bf u, v}) = & \left[{}_{<2}H_1 + {}_{<2}H_2\right]  \psi_{<2}({\bf u}) \psi_{<2}({\bf v})         \equiv    Z \psi_1({\bf u, v})  
\\
W'_{u}(r)  = &
   \left[
  \frac {- u E_1}{2}    + b_1 \sqrt{ \frac {2}{u} }
 + a_1 \sqrt { \frac {u}{2}  }
  \right]  -  Z_1 
\nonumber\\
 W'_{v}(r) = & \left[ 
\frac {- v E_2}{2}    + b_2 \sqrt{ \frac {2}{v} }
 + a_2 \sqrt { \frac {v}{2}  } 
   \right]   - Z_2
   \nonumber
    \end{eqnarray}
}
 \item 
 {
 Model 2
\begin{eqnarray}
\label{38}
H \psi_2({\bf u, v}) =& \left[{}_{<2}H_1 + {}_{>2}H_2\right]  \psi_{<2}({\bf u}) \psi_{>2}({\bf v})   
\equiv  Z \Psi_{<2}({\bf r_1}) \Psi_{>2}({\bf r_2})
\\
W'_{u}(r) = &
   \left[
  \frac {- u E_1}{2}    + b_1 \sqrt{ \frac {2}{u} }
 + a_1 \sqrt { \frac {u}{2}  }
  \right]  -  Z_1 
\nonumber\\
 W'_{v}(r) = & \left[ 
\frac {- u E_2}{2}  + 
  \frac {b_2 u^2}{4} + \frac {a_2 u^3}{8}
  \right] 
         -  Z_2
         \nonumber
  \end{eqnarray}
}
\item 
 {
 Model 3
\begin{eqnarray}
\label{39}
H \psi_3({\bf u, v})& = \left[{}_{>2}H_1 + {}_{<2}H_2\right]  \psi_{>2}({\bf u}) \psi_{<2}({\bf v})   
\equiv  Z \Psi_{>2}({\bf r_1}) \Psi_{<2}({\bf r_2})
\\
W'_{u}(r)  = &
   \left[
\frac {- u E_1}{2}  + 
  \frac {b_1 u^2}{4} + \frac {a_1 u^3}{8}
   \right]  
  -  Z_1 
\nonumber\\
W'_{v}(r) =  &
 \left[ 
\frac {- v E_2}{2}    + b_2 \sqrt{ \frac {2}{v} }
 + a_2 \sqrt { \frac {v}{2}  } 
 \right] 
   - Z_2
   \nonumber
   \end{eqnarray}
}
 \item 
 {
 Model 4
\begin{eqnarray}
\label{40}
H \psi_4({\bf u, v})& = \left[{}_{>2}H_1 + {}_{>2}H_2\right]  \psi_{>2}({\bf u}) \psi_{>2}({\bf v})  
\equiv  Z  \Psi_{>2}({\bf r_1}) \Psi_{>2}({\bf r_2})
\\
W'_{u}(r)  = &
 \left[ 
\frac {- u E_1}{2}  + 
  \frac {b_1 u^2}{4} + \frac {a_1 u^3}{8} 
 \right] 
   -  Z_1 
\nonumber\\
W'_{v}(r) = &
 \left[ 
\frac {- u E_2}{2}  + 
  \frac {b_2 u^2}{4} + \frac {a_2 u^3}{8}
  \right] 
 - Z_2
 \nonumber
       \end{eqnarray}
}
\end{itemize}

At first glance, it seems that 4 models require solving 2 qualitatively different types of problems. However, getting rid of irrationality by introducing a new variable $x' \to x^2$ in one type of problem leads to the solution of a second type problem, which in turn coincides with the problem we considered earlier ($\ref{17}$)  with solution ($\ref{18}$).

Thus, we have obtained that in parabolic coordinates all proposed 4 models  
$(\ref{37}$) - ($\ref{40}$)   is the (quasi) exactly solvable models.

Now let us compare the results obtained with those available in the literature \cite{yeg2}. If we consider just a problem with potential ($\ref{2}$), but without a condition ($\ref{3}$), then we obtain in the 16D  the sum of two independent harmonic oscillators that will be dual to the 9D MICZ-problem with the potential $\cos \theta$:
\begin{eqnarray*}
\frac{ \omega_1^2 u_{s} u_{s} }{2}  + 
\frac{ \omega_2^2 v_{s} v_{s} }{2}  
& \equiv 
- E_1  u_{s} u_{s}  - E_2  v_{s} v_{s} 
\\
\to 
\frac {1} {r}
\left[ 
- \frac {E_1 (r +x_9)} {2} - \frac {E_2  (r - x_9)} {2}   
\right] 
& \equiv 
- \frac {E_1} {2} (1 +\frac {x_9} {r}) - \frac {E_2} {2} (1 - \frac {x_9} {r})  \\
  = -
\frac {E_1+E_2} {2}
-
\frac {E_1-E_2} {2}  \cos \theta  
& \equiv -E_{MICZ} + \frac {\Delta w^2}{4} \cos \theta.   
 \end{eqnarray*}
 
The analog of the 4th order anisotropic potential term for oscillator system \cite{yeg2}  is in our 16D case the sum of 2 harmonic oscillators with identical in value but different in sign potential coefficients for nonlinearity of the 4th order  ($b_1= - b_2$). In the 9-dimensional space, this leads to the desired linear term:

\begin{eqnarray*}
b_1 ( u_{s} u_{s} )^2  + 
b_2 ( v_{s} v_{s} )^2  
& =
b
\left[ 
( u_{s} u_{s} )^2  -
 ( v_{s} v_{s} )^2 
 \right] 
  \\
\to 
\frac {b} {r}
\left[ 
\frac {(r +x_9)^2} {4} - \frac {(r - x_9)^2} {4}   
\right] 
& =
\frac {b} {4r}
\left[ 
4  \ r  \ x_9 
\right] 
\\
=
b \ x_9 
& \equiv b r \cos \theta.   
 \end{eqnarray*}

We note in particular that in our case there is still a new term with a higher degree of nonlinearity than the one considered above.


\section{Conclusion}
Some generalization of 16D oscillator by  anisotropic  and nonlinear inharmonic terms and its dual analog for 9D related MICZ-Kepler systems by generalized version of the Kustaanheimo-Stiefel transformation was shown and analyzed.
The exact analytical solutions of the Schrödinger equation for abovementioned  problems for QES class were discused and given for a few coordinates systems. 
 It is shown that there is a correspondence in particular cases with similar results in lower dimensions.

\bibliographystyle{abbrv}


\appendix


\end{document}